\documentclass[aps,prb,superscriptaddress,preprint]{revtex4-2}

\usepackage{amsmath}
\usepackage{amssymb}
\usepackage{graphicx}
\usepackage{xcolor}

\begin{document}

\title{Detecting Strain Effects due to Nanobubbles in Graphene Mach-Zehnder Interferometers}
\author{Nojoon Myoung}
\affiliation{Department of Physics Education, Chosun University, Gwangju 61452, Republic of Korea}
\author{Taegeun Song}
\affiliation{Department of Data Information Physics, Kongju National University, Gongju 32588, Repulic of Korea}
\author{Hee Chul Park}
\email{corresponding author: hc2725@pknu.ac.kr}
\affiliation{Department of Physics, Pukyong National University, Busan 48547, Republic of Korea}

\date{\today}

\begin{abstract}
    We investigate the effect of elastic strain on a Mach-Zehnder (MZ) interferometer created by graphene p-n junction in quantum Hall regime. We demonstrate that a Gaussian-shaped nanobubble causes detuning of the quantum Hall conductance oscillations across the p-n junction, due to the strain-induced local pseudo-magnetic fields. By performing a machine-learning-based Fourier analysis, we differentiate the nanobubble-induced Fourier component from the conductance oscillations originating from the external magnetic fields. We show that the detuning of the conductance oscillations is due to the altered pathway of quantum Hall interface channels caused by the strain-induced pseudo-magnetic fields. In the presence of the nanobubble, a new Fourier component for a magnetic flux $\Phi_{0}/2$ appears, and the corresponding MZ interferometry indicates that the enclosed area is reduced by half due to the strain-mediated pathway between two quantum Hall interface channels. Our findings suggest the potential of using graphene as a strain sensor for developments in graphene-based device fabrications and measurements technologies.
\end{abstract}

\maketitle

\section{Introduction}

Graphene's high carrier mobility\cite{choi2021high,tan2014electronic,wang2010manipulating,bolotin2008ultrahigh} enables it to rapidly respond to changes in its environment, allowing it to be used as a sensitive detector for a wide range of stimuli, including gases\cite{yuan2013graphene,yavari2012graphene,choi2017gas} and light\cite{van2015ultraviolet}. Additionally, its large surface-to-volume ratio makes it well-suited for sensors that require a large sensing area, such as biosensors\cite{wu2010highly,zeng2015graphene}. Remarkable mechanical properties of graphene also make it suitable for use in pressure\cite{tao2017graphene,tian2015graphene} and strain sensors\cite{wang2014wearable,li2012stretchable} as it is resilient to stress and deformation.

Especially, effects of elastic strain in graphene have been attractive topics both in fundamental and in applied sciences, with its key mechanical properties like high strength $\sim$ 100 GPa\cite{lee2008measurement}, elasticity\cite{cadelano2009nonlinear} up to 20\%, and stiffness with Young's modulus $\sim$ 1 TPa\cite{lee2008measurement}. With these properties, strain engineering of graphene involves applying controlled strain to graphene devices to modify its electronic and transport properties. For example, there have been intensive investigations on strain-induced bandgap engineering of graphene\cite{ni2008uniaxial,sahalianov2019straintronics}, giving us opportunities to exploit graphene for transistors\cite{mcrae2019graphene,myoung2022strain}. 

The convergence of advancements in research has set the stage for uncovering the intricate mechanisms underlying strain engineering graphene\cite{pereira2009strain,guinea2012strain}, bolstered by experimental data confirming the occurrence of psuedo-magnetic fields (PMFs) induced by strain\cite{levy2010strain,nigge2019room,banerjee2020strain,klimov2012electromechanical}. In graphene characterized by a honeycomb lattice structure, the charge carriers possess both sublattice and valley degree of freedom, which can be effectively described in a continuum Dirac model using pseudo-spin\cite{beenakker2008colloquium}. The valley degree of freedom of charge carriers in graphene is influenced by the presence of strain through coupling with an effective gauge field $\vec{A}_{ps}$ which satisfies relation $\vec{\nabla}\times\vec{A}_{ps}=\vec{B}_{ps}$, where $\vec{B}_{ps}$ is PMF. Specifically, it is theoretically expected that an additional Berry phase arises in strained graphene since the valley degree of freedom is manipulated by the strain-induced PMF\cite{myoung2020manipulation}. In this way, precise nanoscale engineering of strain in graphene presents exclusive opportunities to manipulate the transport phenomena in graphene which are correlated to the valley degree of freedom. Reversely, the strain-mediated modulation of transport properties indicate that graphene also has a capability to sense the presence of strain in graphene devices.

In this study, we propose a versatile theoretical approach that enables highly sensitive strain sensing through quantum transport measurements. Our novel graphene strain sensor incorporates a Mach-Zehnder (MZ) interferometry within a p-n junction of quantum Hall graphene, resulting in conductance oscillations as magnetic flux varies. We demonstrate that the MZ interference is exquisitely affected by the presence of local strain near the p-n junction, providing an effective strain-sensing mechanism. For a theoretical model, we discover how our graphene strain sensor works by showing its sensing mechanism based on strain-induced alteration of MZ interference. When a local strain due to lattice deformation like nanobubbles exists near the MZ interferometry, we demonstrate that the emergent PMF can create additional pathways in the middle of the interferometry. Moreover, we design a feasible device architecture for the graphene strain sensor, achieved by controlling quantum Hall (QH) interface channels through a gate-tunable p-n junction steepness. Utilizing a machine-learning signal processing technique, we propose an advanced and efficient methodology for numerical analysis of results from the MZ interference. We also showcase the graphene MZ interferometry's exceptional strain-sensing capability, allowing it to accurately detect the presence of even minute local strains. Remarkably, our strain sensor displays high precision, effectively capturing strain variations across the graphene sample, regardless of their magnitude. This capability makes it particularly valuable for real-world applications where the detection of local strains is crucial.

\section{Model}\label{sec:MZI}

We commence our investigation by introducing the system of interest: the graphene MZ interferometry for strain sensing. This specific configuration involves a p-n junction operating within the QH regime, resulting in the creation of QH interface channels along the p-n junction\cite{klimov2015edge,kumar2018equilibration,lohmann2009four}. Notably, the equilibration of $\nu=1$ QH edge states does not support Mach-Zehnder interferometry\cite{wei2017mach,jo2021quantum,myoung2017conductance}, but higher $\nu$ QH states generate spatially separated QH channels at the junction, forming a finite area enclosed by pathways of these channels. While random valley mixing at the junction interface diminishes quantum interference effects, we deliberately exclude any disorder or impurity responsible for the coherent-to-Ohmic transition\cite{low2009ballistic,myoung2017conductance}, in our study.

\begin{figure}[ht]
    \centering
    \includegraphics[width = \linewidth]{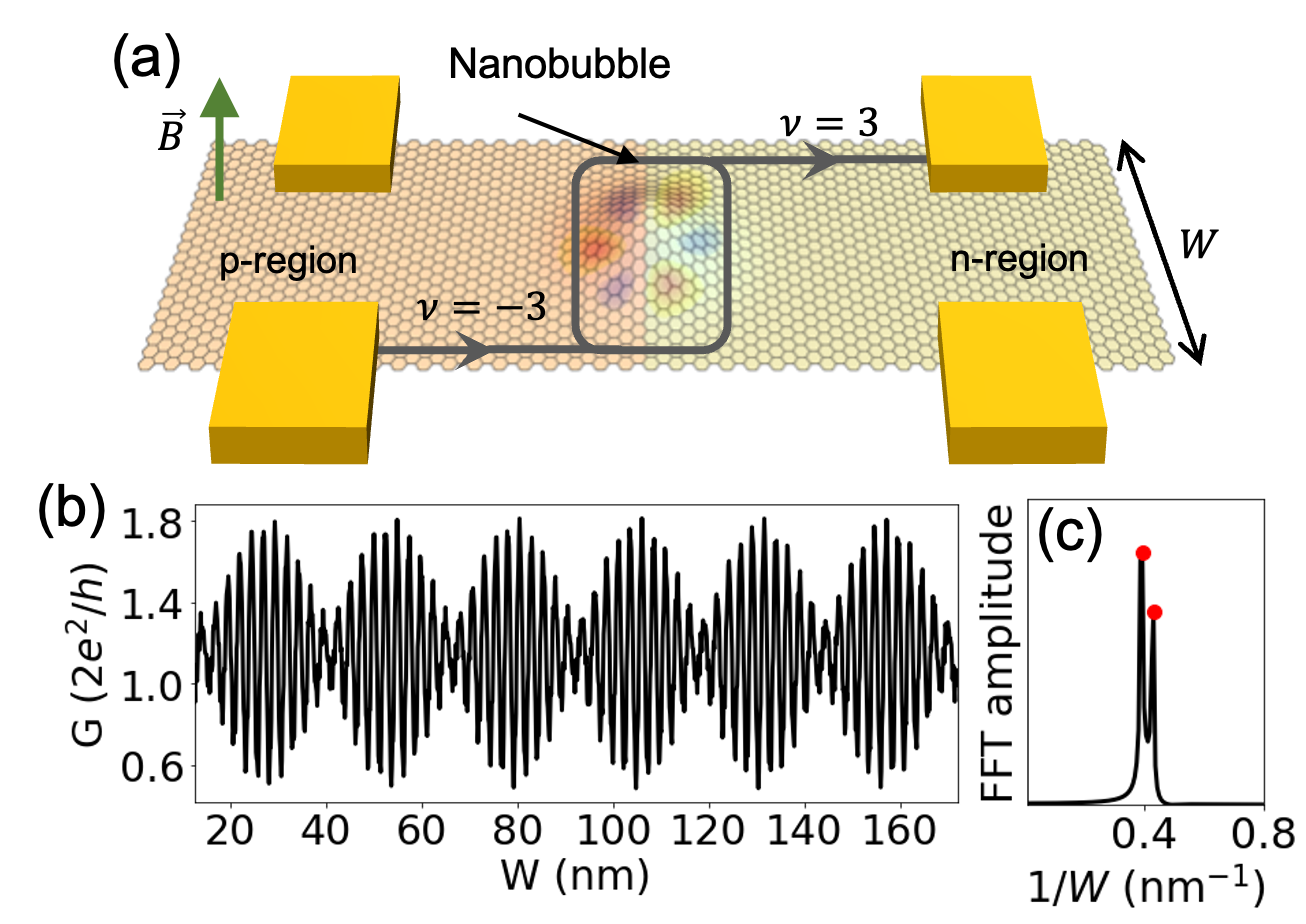}
    \caption{(a) Schematic view of the strain sensor using MZ interferometry with $\nu=3$ QH edge states, capable of detecting local strain near the p-n junction through the sensing of the strain-induced PMF. (b) Conductance oscillation via graphene MZ interferometry as the interface channel width increases, corresponding to the magnetic flux passing through areas enclosed by $\nu=3$ QH channels. (c) FFT spectrum corresponding to the oscillation in (b), with the split FFT peaks denoted by red dots reflecting the valley splitting across the QH channels and closely matching the beating behavior of the conductance oscillation.}
    \label{fig:model}
\end{figure}

Previous reports have indicated that the graphene MZ interferometer generally exhibits low visibility due to asymmetric valley mixing at the p-n junction interface\cite{wei2017mach,myoung2017conductance}. The visibility is further compromised when the p-n junction potential is asymmetrically applied, resulting in unequal filling factors in each region (e.g., $\nu=+1$ in the n region and $\nu=-3$ in the p region). To address this, we focus on a symmetric p-n junction defined by the following potential function along the $y$-direction at $x=0$:
\begin{align}
U\left(x\right)=U_{0}\tanh{\left(\frac{x}{\xi}\right)},
\end{align}
where $U_{0}=\left(1+\sqrt{2}\right)E_{0}/2$, and $\xi$ determines the steepness of the p-n junction. Figure \ref{fig:model}(a) displays the schematic model of graphene MZ interferometry with the channel width $W$, which consists of $\nu=\pm3$ QH interface channels. A larger value of $\xi$ results in a smoother varying potential, while for $\xi\rightarrow0$, the p-n junction potential becomes an abrupt step. The graphene sheet boundaries along the $x$-direction are terminated by armchair configurations, making the MZ interferometer dependent on the valley-isospin at these armchair boundaries\cite{myoung2017conductance,tworzydlo2007valley}. Throughout this paper, we consider a zero angle between the two valley isospins at opposite boundaries, which corresponds to metallic armchair graphene nanoribbons.

In this paper, the transport properties of the system are numerically calculated using the tight-binding approach. As we choose an appropriate gauge $\vec{A}=\left(-By,0,0\right)$ satisfying the homogeneous external magnetic field $\vec{B}=\left(0,0,B\right)=\vec{\nabla}\times\vec{A}$, the tight-binding Hamiltonian can be expressed by
\begin{align}
    H=\sum_{\left<i,j\right>}t_{0}e^{i\varphi_{ij}}a^{\dagger}_{i}b_{j}+\sum_{i}u_{i}\left(a^{\dagger}_{i}a_{j}+b^{\dagger}_{i}b_{j}\right)+\mbox{h.c.},
\end{align}
where $t_{0}=-3$ eV is the nearest hopping energy of graphene, $\varphi_{ij}$ is the phase acquired by Dirac fermions moving through the system via the hopping processes:
\begin{align}
    \varphi_{ij}=-\frac{1}{2}\phi\left(x_{j}-x_{i}\right)\left(y_{i}+y_{j}\right),
\end{align}
with $\phi=\Phi/\Phi_{0}$ is the dimensionless flux normalized by the flux quantum $\Phi_{0}=h/e$, and $u_{i}$ is the on-site potential according to the given p-n junction potential:
\begin{align}
    u_{i}=U_{0}\tanh{\left(\frac{x_{i}}{\xi}\right)}.
\end{align}
In presence of strain, we modify the hopping amplitude for the nearest hoppings:
\begin{align}
    t_{0}\longrightarrow t_{0}e^{-\beta\left(\frac{d_{ij}}{a_{0}}-1\right)},
\end{align}
where $d_{ij}$ is the deformed carbon-carbon distance under a strain and $a_{0}=0.142$ nm is the undeformed distance between adjacent carbon atoms. With the tight-binding Hamiltonian, we carry out numerical calculation of the ballistic conductance across the p-n junction with Landauer formalism,
\begin{align}
    G\left(E\right)=\frac{2e^{2}}{h}\sum_{i\in N}T_{i}\left(E\right),
\end{align}
where $N$ is the number of propagating modes in a incident lead, and $T_{i}$ is the transmission function through $i$-th propagating mode, acquired using the tight-binding S-matrix from \textsc{kwant} codes\cite{groth2014kwant}. In this work, for simplicity, we consider Fermi energy of the system is zero, i.e., $E=0$, so that the $\nu=3$ QH interface channels are symmetrically created at the center of the p-n junction.

The coupling of the strain-induced PMF with the valley degree of freedom in graphene leads to conductance oscillations in the MZ interferometer, arising from valley-isospin rotation\cite{myoung2020manipulation}. In Figure \ref{fig:model}(b), we observe the oscillatory behavior of quantum conductance across the p-n junction without nanobubbles, as the QH channel width varies along the MZ interferometer. The results exhibit a beating nature, indicating the coincidence of two similar frequencies related to valley splitting due to the interplay between external and PMFs\cite{myoung2020manipulation}. The analysis of the results, shown in Figure \ref{fig:model}(c), apparently reveals two peaks in the FFT frequency, $f_{1}=0.395~\mathrm{nm}^{-1}$ and $f_{2}=0.432~\mathrm{nm}^{-1}$, corresponding to a beat period of $1/\left(f_{2}-f_{1}\right)=26.6$ nm. Through a detailed investigation of the beating characteristics of the graphene MZ interference with $\nu=3$ states, we gain insight into the discrimination of valley-resolved pathways\cite{myoung2017conductance}.

Before closing this section, we should mention the presence of second-harmonic MZ interference in the FFT analysis. Although it is not evidently visible in Figure \ref{fig:model}(c) due to its very small amplitude, we find a secondary FFT component at $f_{3}\sim 0.821~\mathrm{nm}^{-1}$, which is approximately double the fundamental MZ frequencies $f_{1}$ and $f_{2}$. Typically, such a weak-amplitude FFT peak becomes more distinguishable with a larger dataset for the FFT processing. Nevertheless, in this work, we successfully identify the secondary effects of the MZ interferometer by employing machine-learning-based Fourier transform techniques. Further details are discussed in the following section.

\section{Results and Discussion}

To discuss the strain sensing capability, we first examine the effects of local strain on the MZ interference. By conducting FFT analysis on the numerical results, we reveal the emergence of new FFT frequency components when a nanobubble exists at the p-n junction, with a frequency that is half that of the fundamental MZ interference. This analysis highlights the division of the enclosed area in the MZ interferometry into two equal areas. Furthermore, we emphasize that the appearance of the emergent MZ pathway is due to the presence of PMF induced by strain in the nanobubble. However, the strain-induced FFT peaks are much smaller than the fundamental MZ interference signals, making it challenging to identify their emergence through the FFT analysis alone. Therefore, to overcome this limitation, we present a machine-learning technique that accurately captures the exact values of frequency components embedded in the beat oscillation spectra, akin to the theoretical Fourier transform.

\subsection{Strain effects on MZ interference}\label{sec:strain}

We carry out the conductance calculation through the graphene MZ interferometry with a nanobubble. The PMF due to the nanobubble is typically characterized by two length parameters, height($h_{0}$) and size($\sigma$) for Gaussian-shaped deformation:
\begin{align}
    z\left(x,y\right)=h_{0}e^{-\left(x^{2}+y^{2}\right)/2\sigma^{2}}.
\end{align}
For simplicity, we set $\sigma=7.38$ nm as a constant, and vary $h_{0}$ from 0 to 4.92 nm to change in the PMF strength. In other words, throughout this paper, the effects of strain are controlled by varying $h_{0}$ values.

\begin{figure}[ht]
    \centering
    \includegraphics[width = \linewidth]{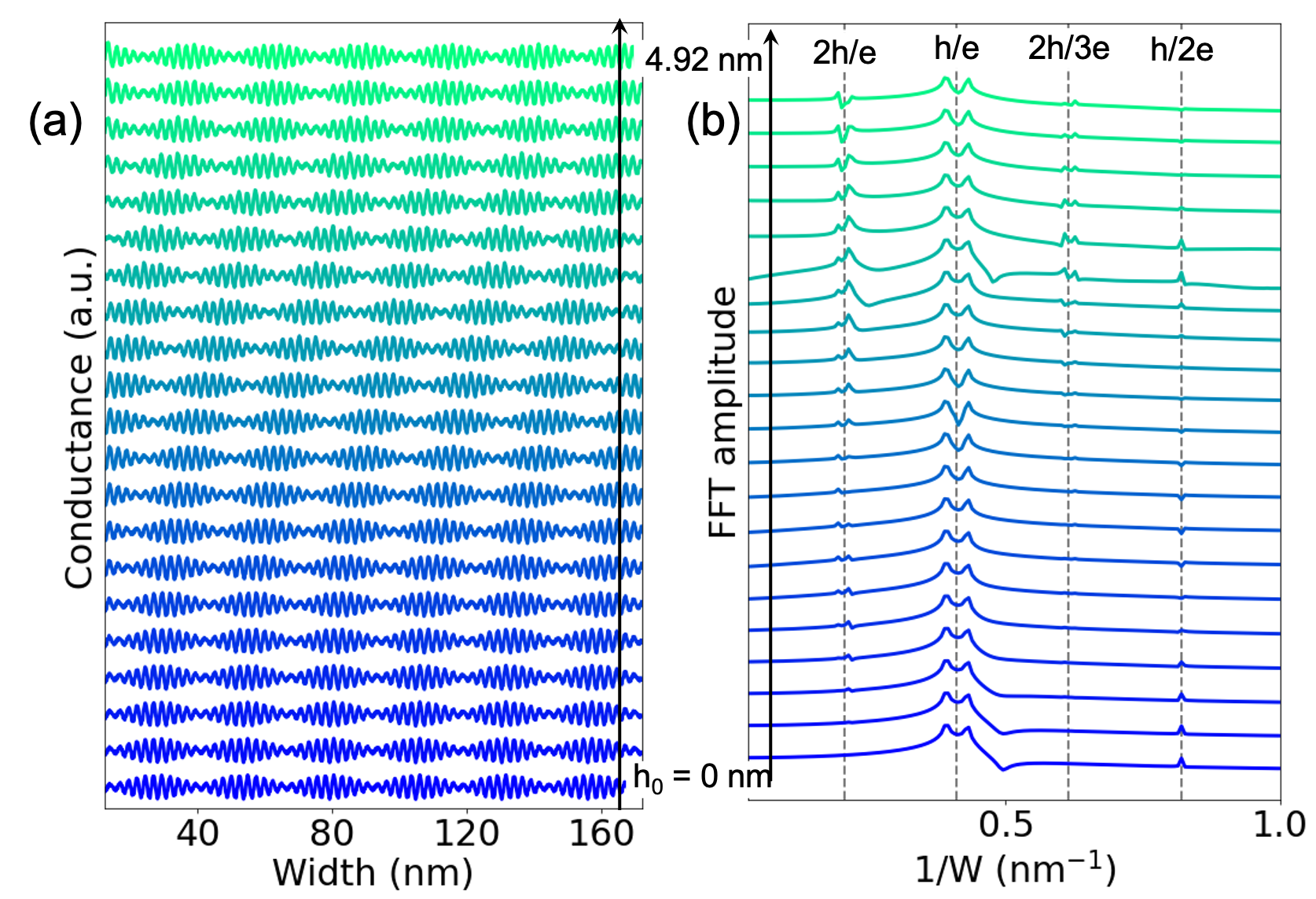}
    \caption{(a) Conductance spectra through the graphene MZ interferometer for various PMF strengths as functions of the QH channel width $W$. The PMF strength due to the nanobubble is varied by $h_{0}$ from 0 nm to 4.92 nm, with the given $\sigma=7.38$ nm. (b) The FFT analysis results of the graphene MZ interference for various PMF strengths. Each FFT spectrum is plotted in logarithmic scale. The dashed lines represent the theoretical estimation of magnetic fluxes piercing the MZ interferometer, $2\phi_{0}$, $\phi_{0}$, $2\phi_{0}/3$, and $\phi_{0}/2$, respectively. $\phi_{0}=h/e$ is the flux quantum for an electron.}
    \label{fig:strain}
\end{figure}

Figure \ref{fig:strain} illustrates the impact of local strain on the graphene MZ interferometer, alongside the FFT analysis of the strain-induced MZ pathways. While the conductance oscillations exhibit slight changes in their curves with increasing PMF strength, their FFT spectra clearly reveal the emergence of peaks resulting from strain effects. In the absence of strain, i.e. $h_{0}=0~\mathrm{nm}$, fundamental and secondary MZ interference peaks are observed, corresponding to magnetic fluxes of $h/e$ and $h/2e$ piercing the enclosed area. It is worth noting that the splitting of these MZ interferences is a result of the valley splitting in the QH interface channels, as reported previously\cite{myoung2017conductance}. The presence of the $h/e$ and $h/2e$ signals can be well interpreted by evaluating the magnetic fluxes through the MZ interferometry $\Phi=BW\Delta x=h/e$ or $h/2e$, respectively. Here, $\Delta x$ indicates the separation of two $\nu=\pm3$ QH channels in both regions, determined by $\Delta x=2r_{c}^{rms}=2r_{c}/\sqrt{2}=2l_{B}$, where $r_{c}^{rms}$ is the root-mean-square value of the snake-like cyclotron orbit at the p-n junction. For the $h/e$ FFT peaks, the conductance oscillations of the MZ interferometer are characterized by the fundamental frequency $k_{f}=2eBl_{B}/2\pi \hbar=1/\pi l_{B}$. In this study, we find $k_{0}=0.411~\mathrm{nm}^{-1}$ with $l_{B}=0.775~\mathrm{nm}$ and $B_{ext}=1097~\mathrm{T}$. From the FFT analysis of the numerical results, the fundamental peaks are found to be $k_{f1}=0.395~\mathrm{nm}^{-1}$ and $k_{f2}=0.432~\mathrm{nm}^{-1}$ as shown in Figure \ref{fig:strain}(b). Their average $\left(k_{f1}+k_{f2}\right)/2=0.414~\mathrm{nm}^{-1}\simeq k_{f}$ is in good agreement with our MZ interferometry. Similarly, the frequency of the secondary conductance oscillation is calculated by considering the MZ interferometry and is found to be $0.822~\mathrm{nm}^{-1}$, which is consistent with the FFT signals at $0.821~\mathrm{nm}^{-1}$ in Figure \ref{fig:strain}(b).

On the other hand, with the presence of the nanobubble, the FFT analysis spectra of the MZ interferences clearly exhibit emergent signals near $2h/e$ and $2h/3e$, in addition to the $h/e$ and $h/2e$ peaks. The appearance of the $2h/e$ signals is particularly intriguing as it indicates that the area enclosed by the MZ pathways is halved, resulting in a magnetic flux through the area equal to the flux quantum. In other words, the non-uniform PMF induced by the nanobubble introduces a new pathway for the MZ interference, effectively dividing the enclosed area into two regions of equal size. For instance, the FFT spectra of the conductance oscillation with $h_{0}=4.97~\mathrm{nm}$ exhibit strain-induced $2e/h$ peaks at $0.194~\mathrm{nm}^{-1}$ and $0.219~\mathrm{nm}^{-1}$, which are approximately half the previously mentioned $h/e$ peaks. Additionally, the FFT spectra also reveal higher-order harmonic generations of the strain-induced MZ interference at $h/e$, $2e/h3$, and so on. Although the second-harmonic generation of the $2e/h$ signals is not easily visible due to its much smaller amplitude compared to the fundamental MZ interference at $h/e$, the third-harmonic generation of the strain-induced MZ interference is still detectable, even though its signals are very weak.

As shown in Figure \ref{fig:strain}(a), an interesting observation regarding the strain effects on the conductance oscillation is the appearance of a phase shift as $h_{0}$ varies. This phase shift arises from the strain-induced Berry's phase acquired by electrons passing through the MZ interferometry. When a nanobubble exists near the p-n junction, the total magnetic flux includes contributions from both the external magnetic fields and the PMF. The beat oscillation of the conductance can be described by the formula:
\begin{align}
G\propto &A_{1}\sin{\left(2\pi\frac{\Phi_{1}+\Phi_{ps,1}}{\Phi_{0}}+\pi\right)}\nonumber\\
&+A_{2}\sin{\left(2\pi\frac{\Phi_{2}+\Phi_{ps,2}}{\Phi_{0}}+\pi\right)}, \label{eq:beatcond}
\end{align}
where $\Phi_{i}$ and $\Phi_{ps,i}$ represent the magnetic fluxes due to the external field and the PMF, respectively, with $i=1,2$ denoting the valley-resolved MZ pathways. The difference between $\Phi_{1}=k_{f1}W$ and $\Phi_{2}=k_{f2}W$ contributes to the beat frequency $k_{f2}-k_{f1}$, while the sum of $\Phi_{ps,1}$ and $\Phi_{ps,2}$ affects the phase shift, given by $\left(\Phi_{ps,1}+\Phi_{ps,2}\right)/\Phi_{0}$. Additionally, the strain-induced phase shift can be further analyzed using complex analysis of the FFT results. Taking the argument of the FFT results as complex numbers allows us to extract information about the phase for each FFT peak frequency, corresponding to the strain-induced Berry's phase $\Phi_{ps}$. It is worth noting that the pristine graphene's Berry's phase $\pi$ is simply added, even though it might not be explicitly reflected in the resulting conductance oscillation curves.

\begin{figure*}[ht]
    \centering
    \includegraphics[width = \linewidth]{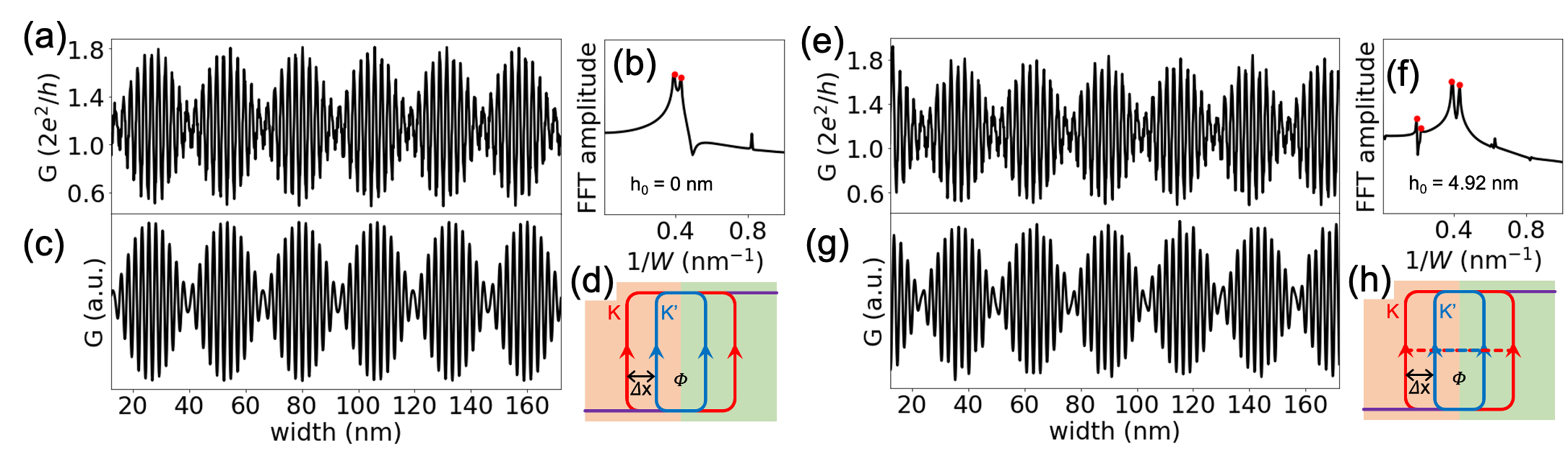}
    \caption{(a) and (e) Conductance oscillations of the graphene MZ interferometer under different conditions: without a bubble ($h_{0}=0~\mathrm{nm}$) and with a bubble ($h_{0}=4.92~\mathrm{nm}$). (b) and (f) FFT analysis of the calculated conductance. The FFT spectra are presented on a logarithmic scale. (c) and (g) Reproduced conductance oscillation curves obtained using Equation (\ref{eq:Fourier}). These conductance reproductions consider the fundamental MZ peaks near $2e/h$ and $e/h$, as indicated by the red dots in the FFT analysis results of (b) and (f). (d) and (f) Schematics illustrating the MZ pathways enclosing the areas through which the external magnetic flux $\Phi$ pierces, both without and with a nanobubble. The red and blue lines represent the valley-resolved MZ pathways for the $\nu\pm 3$ QH channels. The difference $\Delta x$ between the two MZ pathways arises due to the valley splitting at the p-n junction. The dashed lines in (f) depict the strain-induced additional paths, leading to the emergence of half-area MZ interferences.}
    \label{fig:MZpath}
\end{figure*}

To check the effects of strain on the conductance oscillation, we quantitatively reconstruct the MZ interference-based conductance oscillation by plugging the FFT analysis outcomes into the following equation:
\begin{align}
G\left(W\right)\propto \sum_{i}A_{i}\sin{\left(2\pi k_{i}W+\varphi_{i}\right)}, \label{eq:Fourier}
\end{align}
where $A_{i}$, $k_{i}$, and $\varphi_{i}$ represent the amplitude, frequency, and argument of the complex-number FFT data at the $i$-th peak position in the FFT spectra. It is noteworthy that we focus solely on the fundamental generation of half and full-area MZ interferences near the FFT frequencies of $2e/h$ and $e/h$.

In the absence of a nanobubble, the graphene MZ interferometer exhibits two prominent FFT peaks near $e/h$, as depicted in Figure \ref{fig:MZpath}(b), with no signals at $2e/h$. Notably, the reproduced conductance oscillation shown in Figure \ref{fig:MZpath}(c) demonstrates a good agreement with the numerical results depicted in Figure \ref{fig:MZpath}(a). As previously mentioned, this splitting phenomenon is attributed to the valley splitting at the p-n junction, resulting in the separation of the QH channels by an amount of $\Delta x$. The spatial difference between the two valley-resolved MZ pathways, $\Phi_{1}=BW\left(x+\Delta x/2\right)$ and $\Phi_{2}=BW\left(x-\Delta x/2\right)$, is very small ($\Delta x\sim 0.1~\mathrm{nm}$, in this study). Furthermore, our MZ interference mechanism demonstrates a remarkable sensitivity to changes in the QH interface channels, a feature often linked to gate voltage and external magnetic fields.

In the presence of a nanobubble, a different scenario unfolds. As observed in Figure \ref{fig:MZpath}(f), there are now four prominent FFT peaks near $2e/h$ and $e/h$. This observation is further supported by Figures \ref{fig:MZpath}(e) and (g), where the reproduced conductance oscillation through FFT analysis aligns well with the numerical results. As previously discussed, these strain-induced FFT peaks are linked to the emergence of additional MZ pathways, which enclose half the area of the conventional MZ loop, as visualized in Figure \ref{fig:MZpath}(h). In addition, it is worth noting that higher-order generations of the MZ interference appear in the FFT spectra, as depicted in Figures \ref{fig:MZpath}(b) and (f), but their impact remains imperceptible, even when included in Equation (\ref{eq:Fourier}).

\subsection{Accurate Fourier Analysis using Machine Learning}

Here, we present a concise overview of our novel approach to Fourier analysis through machine learning (ML) methodology. To facilitate our purpose, we adopt a structure employing a single hidden-layer artificial neural network, composed of perceptrons\cite{perceptron}. Notably, we focus on the neural network's singular input and output. We refer to values as input $x$, output $y$, and $z$ for the hidden layer. The relationship between these variables and the associated parameters – weights denoted as $\theta_w$ and biases denoted as $\theta_b$ – can be directly expressed as follows: $z=f(\theta_{w0}\cdot x + \theta_{b0})$, and $y=g(\theta_{w1}\cdot z + \theta_{b1})$. Wherein the activation function $f$ connects the input $x$ to the hidden layer, and function $g$ produces the output y from the hidden layer. 
Extending the number of hidden nodes becomes a straightforward process. In this case, the computation rule for the output aggregates the values across all hidden nodes within a single hidden layer: $z_{k}=f\left( \theta_{w0}^{k}\cdot x + \theta_{b0}^{k} \right)$, and $y=g\left( \sum_{k=1}^{M} \theta_{w1}^{k}\cdot z_{k} + \theta_{b1}\right).$ The $k$ denotes the index of the added node runs from one to $M$. By putting the activation functions as the sine function for $f$, and the simple linear function for $g$, one can rewrite the relation as: 
$$y= \sum_{k=1}^{M} \theta_{w1}^{k} \sin \left( \theta_{w0}^{k}\cdot x + \theta_{b0}^{k} \right) +\theta_{b1}$$
The above relation is directly compared with the Equation (\ref{eq:Fourier}). Since the basal idea of the neural net is optimizing network parameters to express output y to the given input x, a well-developed optimization method in a supervised learning manner provides undetermined network parameters. Thus, after succeeded the training procedure, we obtain parameter sets $\{ \theta_{w0}^{k}\}$, $\{\theta_{w1}^{k}\}$, and  $\{\theta_{b0}^{k} \}$ that correspond to mapped into $\{ k_i\}$, $\{A_i\}$, and $\{\phi_{i}\}$ in Equation (\ref{eq:Fourier}), respectively.

To substantiate the previously elucidated method, \textsc{TensorFlow}\cite{tensorflow2015-whitepaper} was utilized to both construct and train the ML kernel. Specifically, the network architecture incorporates a thousand hidden nodes $(M=1000)$, linking connections between the input and output nodes. Employing the Adam optimizer\cite{adam} in conjunction with the mean squared error (MSE) loss function, we effectively honed the ML kernel. This kernel was trained using conductance spectra as functions of the QH channel width while maintaining a given nanobubble height, $h_0$. A clear distinctive advantage of our ML based approach is its ability to access exact frequency information. Furthermore, it has been corroborated that this approach remains effective even when dealing with short signals where FFT analysis falters.

\begin{figure*}[ht]
    \centering
    \includegraphics[width = \linewidth]{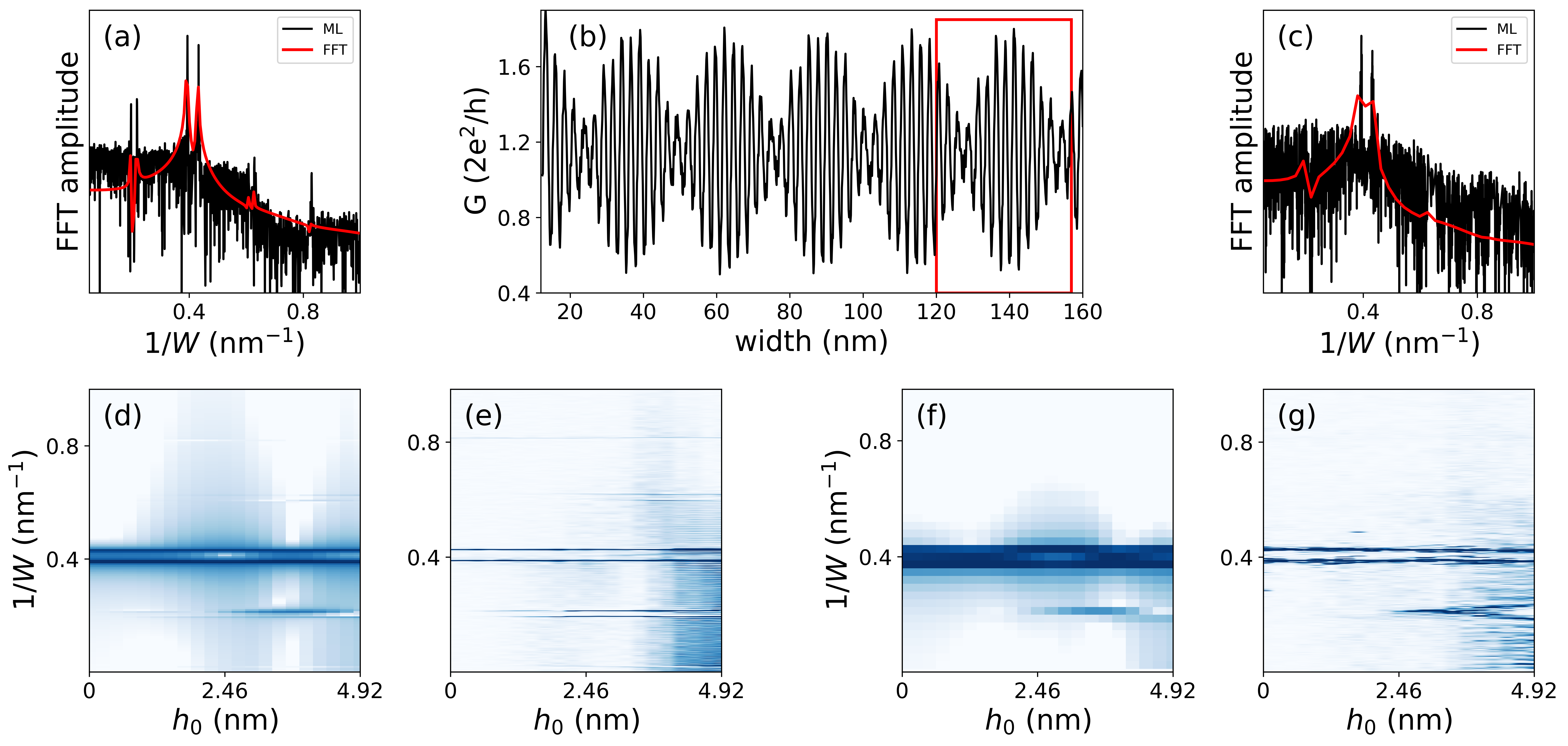}
    \caption{(a) and (c) Comparative analysis of results generated by FFT and ML approaches, highlighting the distinct scales employed by these methods, for $h_0=4.92$ nm. (b) Datasets in use for both FFT and ML approaches. The red rectangle indicates the selected data subset used for the short signals to be analyzed in (c). (d)-(g) Results of Fourier analysis for varying bubble heights using the FFT and ML approaches. For (d) and (e), the complete dataset shown in (b) is taken into account Fourier analysis, whereas for (f) and (g), the data subset highlighted in (b) is utilized for the short-signal analysis. Identical datasets were utilized for pairs (d) and (e), (f) and (g). The color maps of (d) and (f) are presented on a logarithmic scale for FFT results in (d) and (f), while a linear scale is employed for ML results in (e) and (g). Note that while the FFT analysis only slightly reveals the presence of $2\phi_0 /3$ and $2\phi_0$ MZ signals in a logarithmic scale, these signals are clearly observed using the ML approach, even when represented on a linear scale. }
    \label{fig:MLcomp}
\end{figure*}

The performance of the trained ML kernel is illustrated in Figure \ref{fig:MLcomp}. In Figure \ref{fig:MLcomp}(a), we plot the results of FFT analysis and the ML approach. We conducted individual training sessions for $10$ ML kernels, each comprising $1000$ hidden nodes. It is noteworthy to note that we concluded the training phase as the MSE value approached approximately $\mathcal{O}(10^{- 5\sim 6})$. Subsequently, we collect the network parameter values $\{\theta_{w0}^{k}\}$, and $\{\theta_{w1}^{k}\}$ to estimate the frequencies of oscillating conductances. This process involved segmenting the frequency range by the mean of MSE, followed by summating the amplitudes, $\theta_{w1}^{k}$ within these segments, and averaging across the trained 10 ML kernels. The ML results exhibited an impeccable concordance with the FFT analysis, as depicted in the plot. These results encompass not only fundamental signals but also intricate MZ interference signals, situated near $2h/e$ and $2h/3e$. In addition, with the FFT method, the short signal noted by the red rectangle box in Figure \ref{fig:MLcomp}(b) is indistinguishable two fundamental frequencies $k_{f1}$ and $k_{f2}$. However, remarkably, the ML approach makes it available as shown in Figure \ref{fig:MLcomp}(c).

The results depicted in Figure \ref{fig:MLcomp}(d)-(g) illustrate summarized color maps that encompass the entire dataset. Panels (d) and (e) utilize the entirety of the data displayed in (b), while panels (f) and (g) are used short input signals as shown a red box in (b). Upon considering the comprehensive dataset, the presence of two fundamental peaks and strain-induced emerging peaks are clearly confirmed. The application of ML analysis yields notable outcomes, revealing frequency peaks characterized by a width of MSE, thereby suggesting a demand for extensive time-domain data (order of width window to be $\approx 10^{5}$) for achieving such attributes through FFT analysis. Notably, in shorter time scales, the utilization of machine learning demonstrates clear advantages. These findings underscore the competence of ML methodologies in deciphering intricate patterns of transformation.

\section{Conclusions}

By proposing an innovative method for detecting local strain in graphene, our research outcomes hold significant implications for the potential application of graphene in sensing. The utilization of a graphene Mach-Zehnder (MZ) interferometer with $\nu=\pm3$ quantum Hall channels allows for the detection of changes in conductance oscillations. We have confirmed that the strain-induced additional pathways for Dirac fermions lead to half-area MZ interference. Our observation of strain-induced FFT peak signals at $2e/h$ demonstrates the capability to identify the presence of a nanobubble near the MZ interferometer. Moreover, we have demonstrated the effectiveness of machine-learning Fourier analysis in accurately deciphering weak FFT signals, even when datasets of conductance oscillations are limited for FFT analysis. In conclusion, we infer that our findings hold the potential to initiate the development of highly sensitive graphene strain sensors, leveraging subtle changes in MZ interference patterns.

\begin{acknowledgements}
This research received support from the National Research Foundation (NRF-2022R1F1A1065365, NRF-2022R1F1A1074045) and a research grant of the Kongju National University in 2022. Nojoon Myoung and Taegeun Song contributed equally to this work.
\end{acknowledgements}

\section*{Conflic of Interests}
The authors declare no conflict of interest.

\section*{Data Availability Statement}
The data that support the findings of this study are available from the corresponding author upon reasonable request.

\section*{Keywords}
Graphene p-n junction, Pseudo-magnetic field, Strain sensor, Mach-Zenher interferometer, Machine learning

\bibliographystyle{apsrev4-2}
\bibliography{GraSensorMZ}

\end{document}